\documentclass[manuscript,nonacm]{acmart}

\renewcommand\footnotetextcopyrightpermission[1]{}
\usepackage{graphicx}
\usepackage{booktabs}
\usepackage{xcolor}
\usepackage{listings}
\usepackage{hyperref}
\usepackage{subcaption}

\usepackage{tikz}
\usetikzlibrary{positioning,fit}

\lstdefinestyle{nwqcode}{
  basicstyle=\ttfamily\small,
  columns=fullflexible,
  breaklines=true,
  frame=single,
  numbers=left,
  numberstyle=\tiny,
  xleftmargin=2em,
  framexleftmargin=1.5em
}
\title{NWQWorkflow: The Northwest Quantum Workflow}

\author{Ang Li}
\authornote{This whitepaper reflects the author’s own perspectives on the NWQ software ecosystem and does not represent an official position of PNNL, UW, or the U.S. Department of Energy. The author is transitioning to a new position and this whitepaper serves as a handover summary. The NWQ ecosystem is contributed by many colleagues and supported by several projects, please refer to the acknowledgment section for details. Ang Li serves as the chief computer scientist in the Physical and Computational Sciences Directorate (PCSD), Pacific Northwest National Laboratory, and dual-appointment Associate Professor in the Electrical and Computer Engineering Department (ECE), University of Washington (UW).}
\affiliation{
  \institution{Pacific Northwest National Laboratory (PNNL)}
  \city{Richland}
  \state{WA}
  \country{USA}
}
\email{ang.li@pnnl.gov}
\affiliation{
\institution{University of Washington}
  \city{Seattle}
  \state{WA}
  \country{USA}
}
\email{angli16@uw.edu}

\begin{abstract}
This whitepaper presents NWQWorkflow, an end-to-end workflow for quantum application development, compilation, error correction, benchmarking, numerical simulation, control, and execution on a prototype superconducting testbed. NWQWorkflow integrates NWQStudio (programming GUI environment), NWQASM (intermediate representation), 
QASMTrans (compiler), NWQEC (quantum error correction), QASMBench (benchmarking and characterization),
NWQSim (HPC simulation), NWQLib (algorithm library), NWQData (data sets), NWQControl (quantum control),
and NWQSC (superconducting testbed). The system enables closed-loop software–hardware co-design and reflects the past eight years of quantum computing research the author has led at PNNL (2018–2026). By releasing most software components as open source or planning their open-source availability, we aim to cultivate a collaborative quantum information science (QIS) ecosystem and support the transition toward a scalable quantum supercomputing era.

\end{abstract}

\begin{document}
\maketitle

\section{Introduction}
Over the past decade, significant progress has been made in quantum computing software frameworks and tool development. However, the current ecosystem remains fragmented and presents several limitations for large-scale, scientific quantum computing workloads. This is particularly the case for those relevant to U.S. Department of Energy (DOE) national laboratories and high-performance computing (HPC) facilities.

First, \emph{openness and sustainability remain challenges}. Many widely used quantum software platforms are commercial or vendor-driven (e.g., Qiskit, t|ket⟩, Q\#, Cirq, and PennyLane). While some of these tools are (partially) open source, they are often tightly coupled to specific hardware backends or application domains. Frequent updates and evolving interfaces can introduce broken dependencies and impose a substantial maintenance burden on developers and users seeking long-term stability and interoperability.

Second, \emph{integration across the software stack is limited}. Much of the existing community-developed software is produced in the context of individual research efforts or standalone publications. These tools are typically designed to address individual problems and are not integrated into a cohesive, reusable and systematic infrastructure. As a result, assembling an end-to-end workflow often requires significant manual effort and ad-hoc interfaces, limiting reproducibility and reuse.

Third, \emph{usability and deployability pose barriers}, particularly for quantum HPC environments. Many current quantum software tools are not self-contained and rely heavily on Python-based ecosystems with extensive external dependencies. While suitable for rapid prototyping, such designs can introduce performance overheads and complicate deployment, maintenance, and scaling on shared HPC systems. Moreover, most tools are not explicitly designed with quantum–HPC integration and scaling in mind.

To address these challenges, this whitepaper introduces \textbf{NWQWorkflow}, an end-to-end, open-source, and community-oriented quantum computing ecosystem designed to support scientific quantum applications. The overarching goal of NWQWorkflow is to provide a sustainable and extensible software–hardware workflow maintained by the QIS community, particularly the DOE national laboratory complex.

NWQWorkflow is distinguished by several key features. First, it is \textbf{\emph{open}} by design: most components are open source or on a clear path toward open-source release, and are grounded in research publications. Second, it is \textbf{\emph{comprehensive}}, spanning nearly the entire quantum computing workflow, including programming environments, algorithm libraries, intermediate representations, compilation, quantum error correction, benchmarking, large-scale simulation, data management, control software, and hardware testbeds. Third, the system emphasizes self-containment and performance, with core components, such as the compiler, simulator, quantum error correction framework, and control-stack implemented primarily in C++, and designed to minimize external dependencies. Finally, NWQWorkflow remains device- and application-\textbf{\emph{agnostic}}, targeting general quantum computing workloads without assuming specific hardware modalities or application domains.

Taken together, these design principles position NWQWorkflow as a viable and extensible solution for future quantum computing user facilities and quantum data centers, enabling scalable experimentation and closed-loop software–hardware co-design.

\section{NWQWorkflow Architecture Overview}

\begin{table}[h]
\centering
\caption{Summary of NWQWorkflow components, implementation status, and availability. 'GUI env.' stands for graphic user interface programming environment. 'IR' refers to intermediate representation. 'Alg.' stands for Algorithm. 'HW' stands for hardware. 'mgmt.' stands for management.}
\label{tab:nwqflow-components}
\begin{tabular}{lllp{2.3cm}p{5.1cm}}
\toprule
\textbf{Name} & \textbf{Component} & \textbf{Implementation} & \textbf{Status} & \textbf{Access} \\
\midrule
NWQStudio   & GUI env.                     & Python                    & Ready to release        & Planned: (github.com/pnnl/nwqstudio) \\
NWQASM      & IR         & OpenQASM                  & In progress           & Planned: (github.com/pnnl/nwqasm) \\
QASMTrans   & Compiler                            & C++/Python              & Released              & \url{https://github.com/pnnl/qasmtrans} \\
NWQEC       & QEC compiler                        & C++/Python              & Released              & \url{https://github.com/pnnl/nwqec} \\
QASMBench   & Benchmarking                  & OpenQASM                  & Released              & \url{https://github.com/pnnl/qasmbench} \\
NWQSim      & HPC simulation            & C++/Python & Released              & \url{https://github.com/pnnl/nwq-sim} \\
NWQLib      & Alg. library     & Qiskit                    & In progress           & Planned: (github.com/pnnl/nwqlib) \\
NWQData     & Data mgmt.                     & Text                      & In progress  & Detailed later \\
NWQControl  & Control stack        & C++/Python              & Ready to release        & Planned: (github.com/pnnl/nwqcontrol) \\
NWQSC       & HW testbed      & N/A                       & N/A                   & N/A \\
\bottomrule
\end{tabular}
\end{table}

\begin{figure}[h]
  \centering
\includegraphics[width=0.75\linewidth]{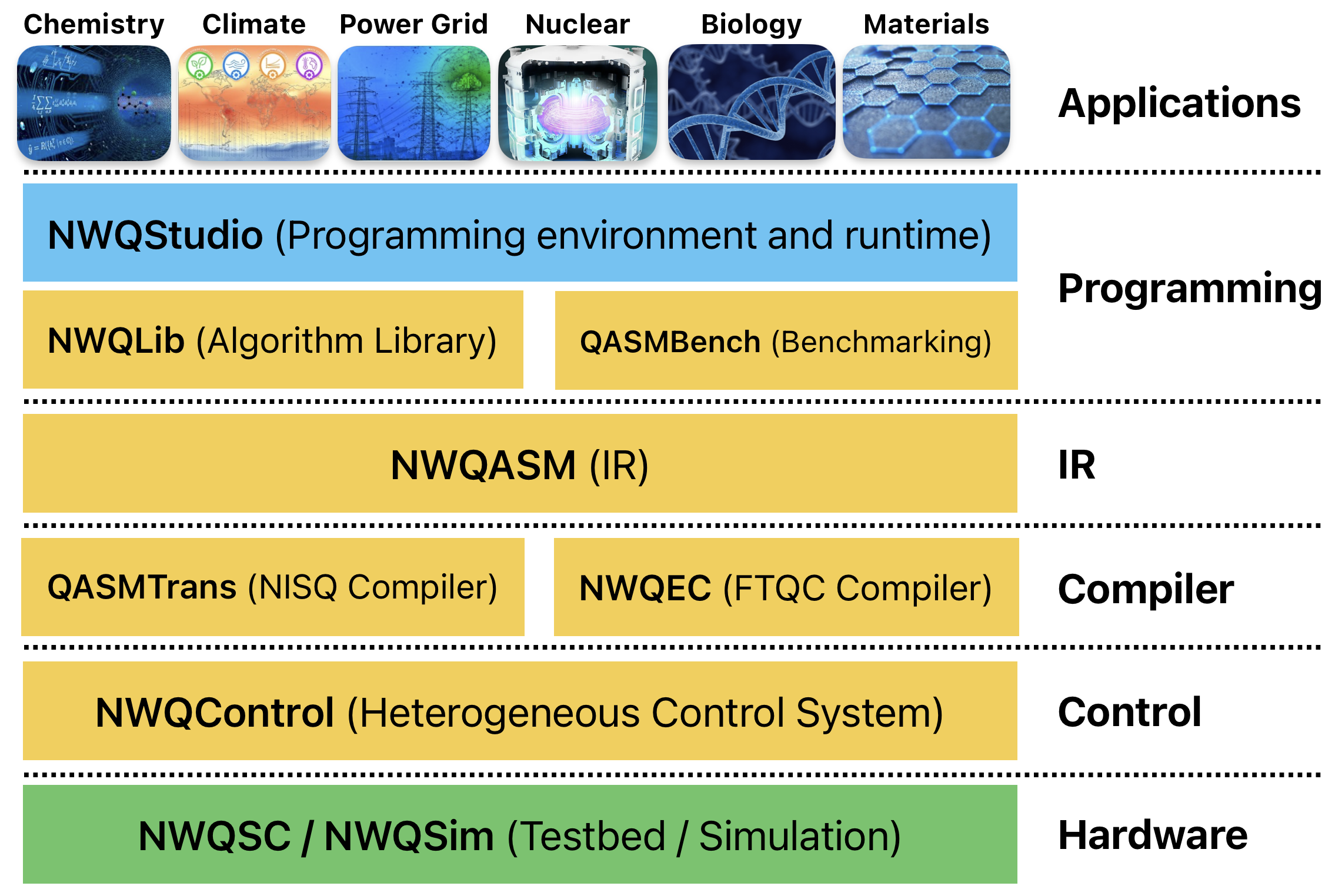}
  \caption{Layered software and hardware stack of NWQWorkflow.}
  \label{fig:nwqflow-stack}
\end{figure}

Table~\ref{tab:nwqflow-components} summarizes the components of the NWQWokflow. The core components have already been released and are publicly available through GitHub. Collectively, these components form a complete ecosystem that spans the entire quantum software stack, presented in the following.

\subsection{System Stack}
Figure~\ref{fig:nwqflow-stack} illustrates the overall NWQWorkflow architecture and the organization of its components within a layered software stack. The workflow is designed to support scientific, mission-driven application domains relevant to the DOE and the national laboratory complex, with particular emphasis on PNNL. While these application domains motivate the design of the system, they are not the focus of this whitepaper. The following sections of this whitepaper describe each component of the stack, proceeding from the programming layer down to the superconducting hardware testbed.

\subsection{Workflow Toolchain Example}

\begin{figure}[h]
  \centering
\includegraphics[width=0.95\linewidth]{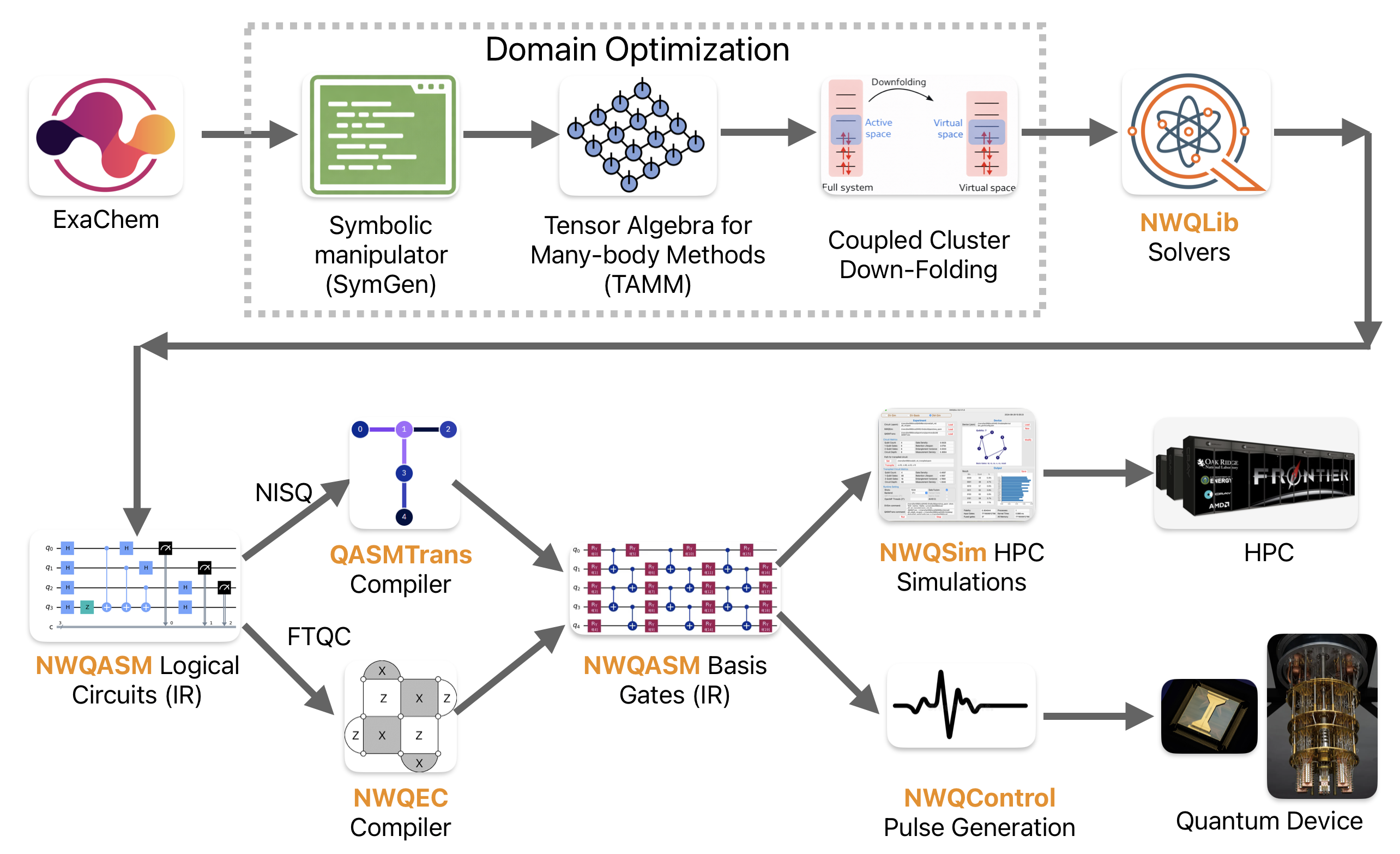}
  \caption{Exemplar quantum chemistry workflow toolchain based on NWQWorkflow.}
  \label{fig:nwqflowchem}
\end{figure}

Figure~\ref{fig:nwqflowchem} illustrates an exemplar quantum chemistry workflow developed at PNNL through close co-design between the chemistry and computing teams. The workflow begins with \textbf{ExaChem}~\cite{panyala2023exachem} (\url{https://github.com/ExaChem/exachem}), a PNNL-maintained suite of scalable electronic structure methods for ground- and excited-state molecular calculations. ExaChem currently provides native implementations of Hartree–Fock (HF), MP2, CC2, CCSD, CCSD(T), CCSD-Lambda, EOM-CCSD, RT-EOM-CCSD, GFCCSD, and double unitary coupled-cluster (DUCC) methods.

The resulting electronic structure Hamiltonians are subsequently processed and optimized using domain-specific techniques, including \textbf{SymGen}\cite{bylaska2024electronic, bauman2025integrated}, \textbf{TAMM}\cite{mutlu2023tamm}, and \textbf{downfolding}~\cite{kowalski2020sub, bauman2022coupled, bauman2025coupled}. \textbf{SymGen} (\url{https://github.com/npbauman/SymGen}) is a software framework for generating symmetry-adapted electronic structure Hamiltonians for quantum simulation of molecular and materials systems. By explicitly exploiting physical symmetries, such as particle number, spin, and spatial symmetries, SymGen reduces problem dimensionality and improves the efficiency of downstream quantum algorithms. \textbf{TAMM} (\url{https://github.com/NWChemEx/TAMM}) is a parallel tensor algebra library that enables performance-portable development of scalable electronic structure methods on HPC platforms.

\textbf{Downfolding} is a systematic model-reduction technique in electronic structure theory that derives effective Hamiltonians capturing the low-energy physics of complex systems while reducing dimensionality for efficient calculation. This approach partitions the full Hilbert space into an active low-energy subspace and an inactive high-energy subspace, with the effects of the latter incorporated into renormalized interactions within the former. As a result, downfolding enables accurate treatment of strongly correlated systems using a substantially reduced number of degrees of freedom. In this workflow, downfolding techniques are implemented and integrated within ExaChem, and selected downfolded Hamiltonians, such as benzene and free-base porphyrin (FBP), are distributed through the DUCC-Hamiltonian-Library~\cite{bauman2025coupled} (\url{https://github.com/npbauman/DUCC-Hamiltonian-Library}).

To compute ground- or excited-state energies from the processed Hamiltonians, quantum algorithm solvers such as ADAPT-VQE~\cite{grimsley2019adaptive} or the GCM~\cite{zheng2023quantum, zheng2024unleashed} method from the \textbf{NWQLib} library are employed. NWQLib is described in detail in a later section. By integrating the selected quantum solver with the corresponding state-preparation routines and ansatz constructions, logical quantum circuits are generated and expressed in the \textbf{NWQASM} format, an extension of the OpenQASM~2.0 intermediate representation (IR)~\cite{cross2017open}. The NWQASM format is introduced in detail later.

These logical NWQASM circuits are subsequently transpiled by our C++-implemented quantum compilers under either a noise-intermediate-scale quantum (NISQ) or fault-tolerant quantum computing (FTQC) execution scenario. In the NISQ setting, circuits are transpiled using \textbf{QASMTrans}~\cite{hua2023qasmtrans}, while in the FTQC setting, compilation and encoding are performed using \textbf{NWQEC}~\cite{wang2024optimizing, wang2025tableau}; both components are introduced in subsequent sections. The resulting physical circuits, still represented in NWQASM but restricted to target basis gates, are then executed either via large-scale numerical simulation using \textbf{NWQSim}~\cite{li2021sv, li2020density} on HPC resources or on physical quantum hardware, such as the \textbf{NWQSC} superconducting testbed, through pulse generation and optimization provided by \textbf{NWQControl}.

\section{NWQWorkflow Components}
The components of NWQWorkflow are presented in a top-down manner through the software stack in Figure~\ref{fig:nwqflow-stack}.

\subsection{NWQStudio IDE: Programming and Evaluation Environment}

\begin{figure}[h]
  \centering

  % ---- Row 1 ----
  \begin{subfigure}[t]{0.495\linewidth}
    \centering
    \includegraphics[width=\linewidth]{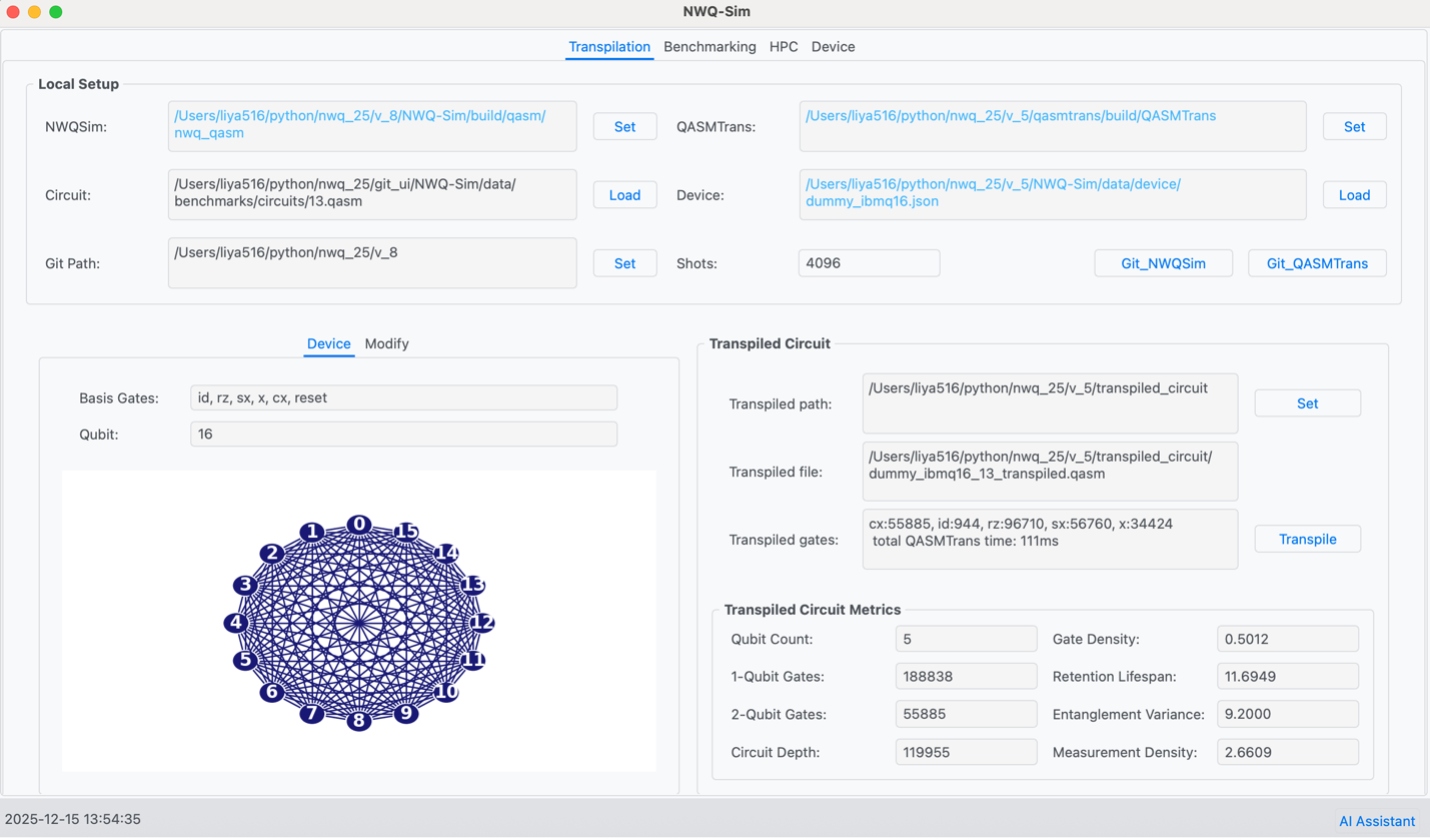}
    \caption{Transpilation tab view.}
    \label{fig:sub_a}
  \end{subfigure}
  \hfill
  \begin{subfigure}[t]{0.495\linewidth}
    \centering
    \includegraphics[width=\linewidth]{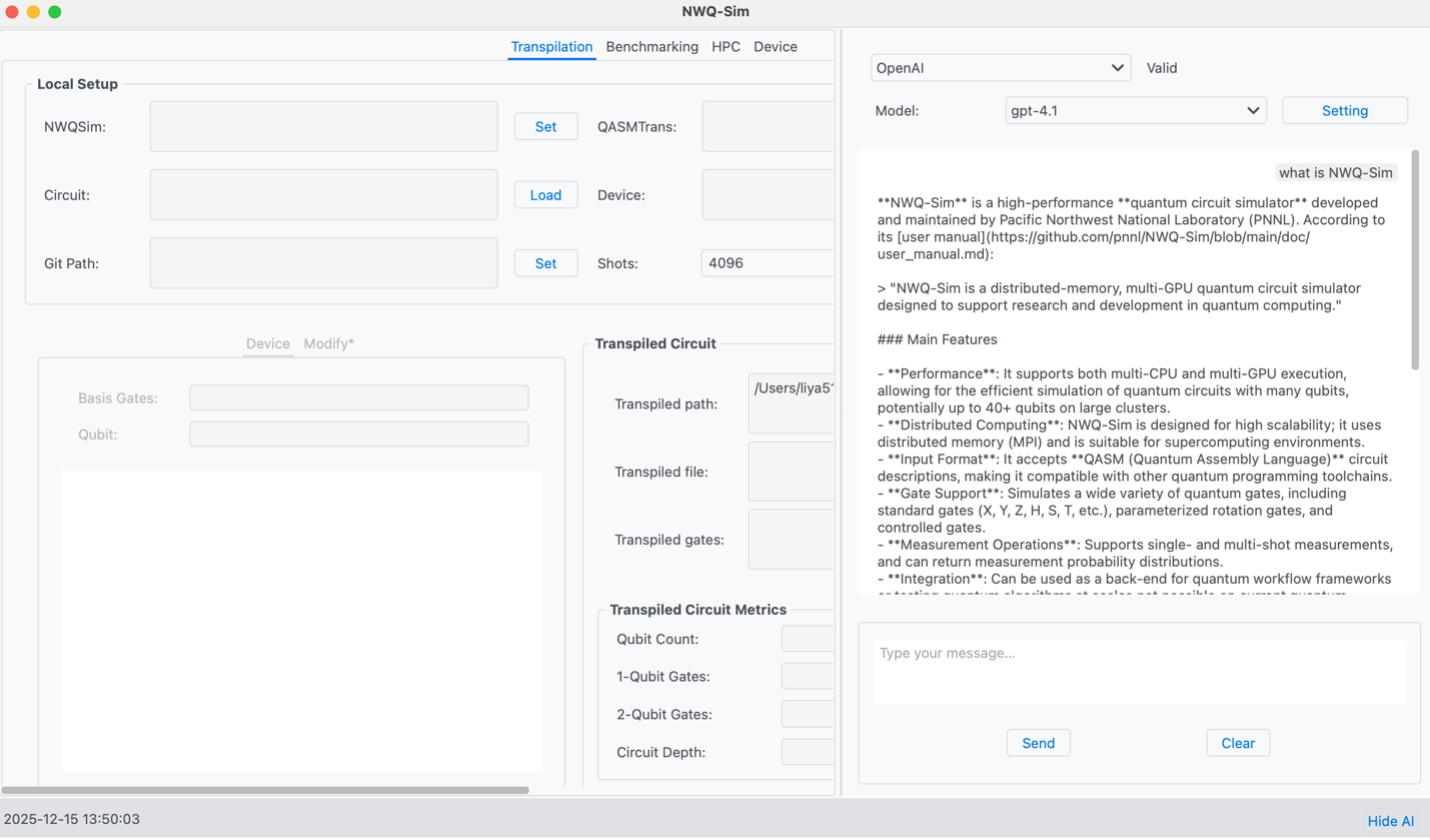}
    \caption{AI agent tab view.}
    \label{fig:sub_b}
  \end{subfigure}

  \vspace{0.8em}

  % ---- Row 2 ----
  \begin{subfigure}[t]{0.495\linewidth}
    \centering
    \includegraphics[width=\linewidth]{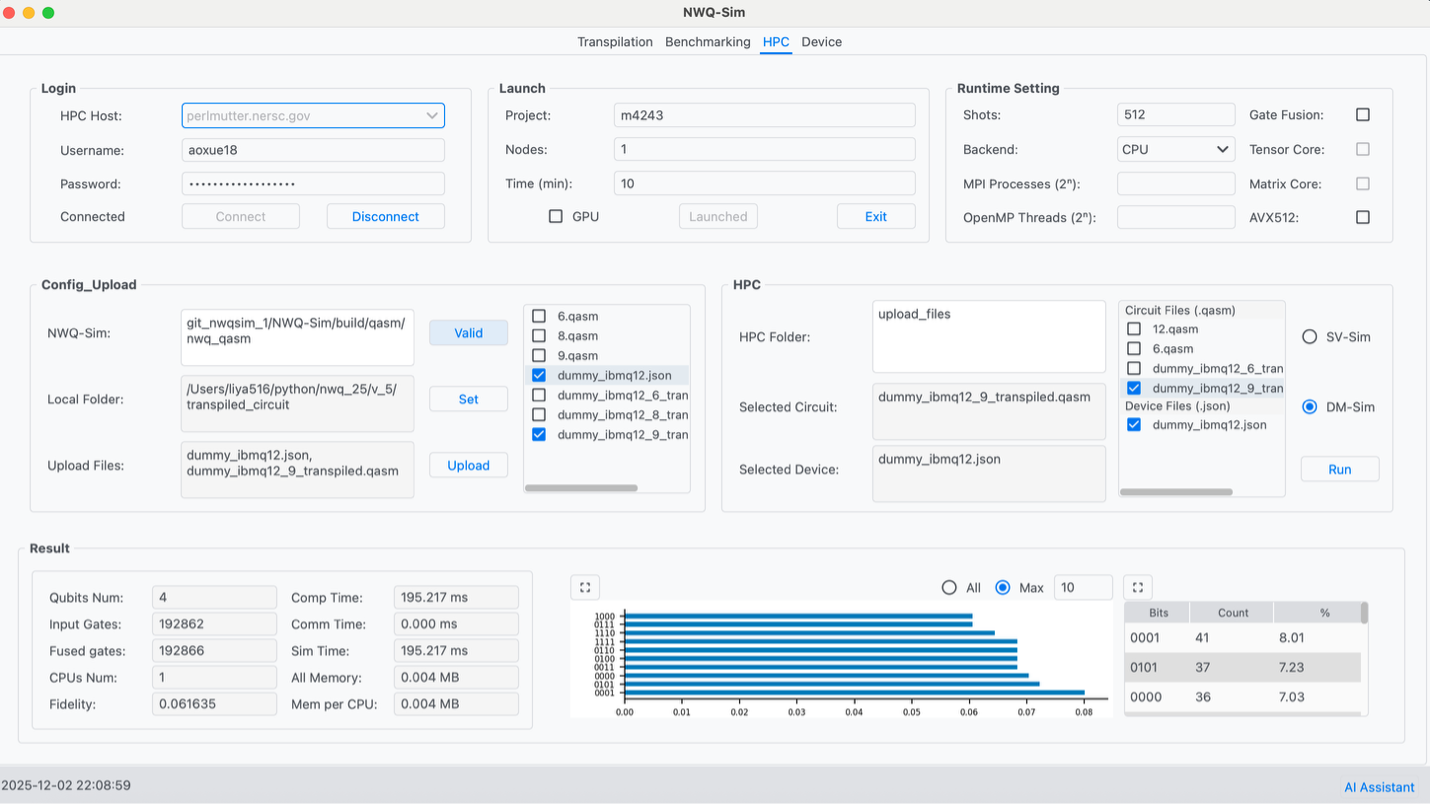}
    \caption{Simulation tab view.}
    \label{fig:sub_c}
  \end{subfigure}
  \hfill
  \begin{subfigure}[t]{0.495\linewidth}
    \centering
    \includegraphics[width=\linewidth]{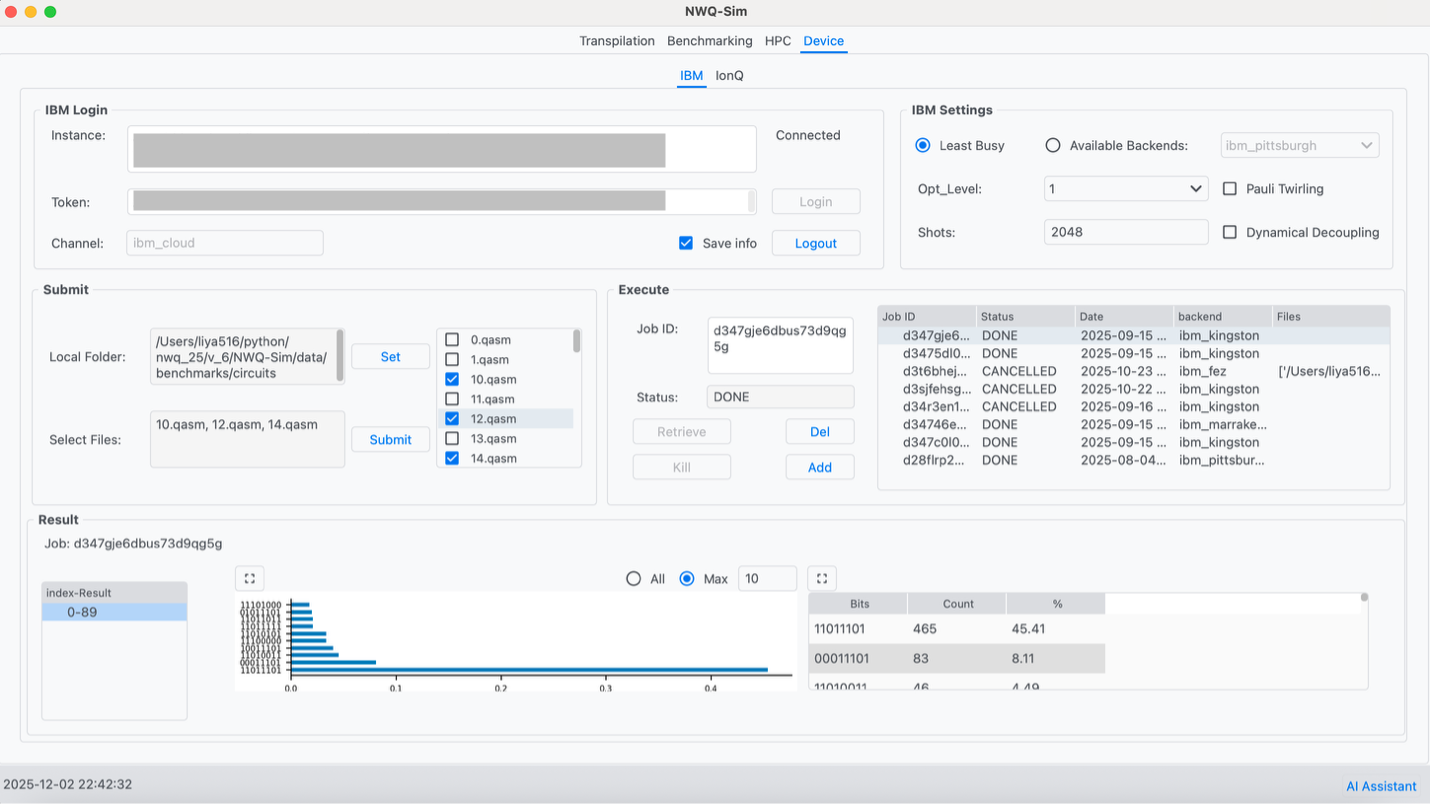}
    \caption{Device execution tab view.}
    \label{fig:sub_d}
  \end{subfigure}

  \caption{NWQStudio Programming and Evaluation Environment.}
  \label{fig:nwqstudio}
\end{figure}

NWQStudio integrates all major components of the NWQWorkflow, including the compiler, simulator, benchmarking, control, and testbed, through a graphical user interface (GUI), forming a comprehensive quantum programming and evaluation environment. The NWQStudio IDE (Integrated Development Environment) is developed in Python using PyQt5 (\url{https://pypi.org/project/PyQt5/}). The original goal was to build a GUI to simplify the usage of the NWQSim simulator on HPC systems. Over time, transpilation, device execution, benchmarking, and AI assistance have been progressively integrated to form a comprehensive IDE. Figure~\ref{fig:nwqstudio} shows exemplar tab views of the NWQStudio GUI.

In Figure~\ref{fig:nwqstudio}-(a), the top panel allows users to configure the input circuit based on QASM or NWQASM IR, specify the paths to the QASMTrans compiler and NWQSim simulator, load the device configuration file (detailed in the following subsection), and set the current workspace. Buttons are provided for automatic Git checkout, compilation, and setup of the QASMTrans compiler and NWQSim simulator. Once the device configuration file is loaded, the bottom-left panel displays the device basis gate set, the number of physical qubits, and a visualization of the device qubit topology. After all configurations are completed and the \textit{Transpile} button is clicked, a transpiled circuit corresponding to the input circuit and device configuration is generated in the specified workspace through QASMTrans. Compilation statistics are also displayed, including the number of different basis gates, 1-/2-qubit gates, circuit depth, gate density, retention lifespan, entanglement variance, and measurement density. The definitions and details of these metrics can be found in our QASMBench paper~\cite{li2023qasmbench}.

Figure~\ref{fig:nwqstudio}-(b) shows the AI agent tab view, where a dedicated frame is used to provide AI assistance within NWQStudio. A separate window is used to configure and test the AI agent, including the LLM API key and model selection (e.g., OpenAI ChatGPT-4.1 here). The configuration supports both direct API key input and linkage to the PNNL AI Incubator local server for access to more advanced models. The original goal of the AI agent was to help users better configure and utilize the tools integrated in NWQStudio, such as setting Slurm parameters for simulations on particular HPC systems, or runtime parameters for launching jobs on real quantum devices. Subsequently, we realized that the AI agent can also assist in reasoning about simulation and execution results, providing feedback for new experiment (e.g., circuit, device, tool) development, and managing NWQStudio jobs, given that simulations on HPCs and executions on cloud-based quantum devices can be time-consuming. The AI agent will further contribute to the objective of building a digital twin through noise-aware density matrix simulations of targeted real devices by automatically launching characterization circuits and using the results to calibrate noise model configuration parameters. This capability enables the construction of an adaptive and accurate noise-aware digital twin for experimental quantum devices, which is valuable for quantum physicists and engineers to deeply investigate device behavior and guide further improvements. Additionally, once the superconducting testbed NWQSC is integrated into NWQStudio, the AI agent will assist with auto-calibration~\cite{fang2025caliqec}, i.e., tracking and compensating for drift in qubit frequencies~\cite{baheri2022quantum}, gate parameters~\cite{stein2022eqc}, readout fidelity~\cite{zheng2023bayesian}, and crosstalk~\cite{hua2022synergistic, kumar2025context}, in order to maintain stable, high-fidelity operations over time.

Figure~\ref{fig:nwqstudio}-(c) shows how to upload and evaluate a (transpiled) quantum circuit using NWQSim-based numerical simulation on an HPC cluster. The figure illustrates the simulation of a QASMTrans-transpiled circuit using DM-Sim, the density matrix simulator of NWQSim, on the NERSC Perlmutter HPC system. Simulation results and runtime statistics are displayed at the bottom of the page, including fidelity, simulation time, and memory usage. In particular, the fidelity metric is highlighted, which is computed by comparing the density-matrix-based noisy simulation (using physical parameters from the device configuration file, such as T1, T2, gate time, and readout fidelity) against an ideal state-vector simulation without noise. The measured bit-string results for a specified number of shots are visualized in a histogram, ranked by their occurrence frequency.

Finally, Figure~\ref{fig:nwqstudio}-(d) shows the evaluation workflow on real quantum devices, such as those provided by IBM and IonQ (with \emph{ibm\_pittsburgh} shown as an example). This page enables runtime configuration, including the IBM instance, authentication token, channel, backend, optimization level, number of shots, and the selection of error mitigation techniques such as Pauli twirling and dynamical decoupling. NWQStudio leverages multi-threading to manage outstanding jobs, where a dedicated thread is spawned to monitor submitted jobs until completion. Job information is listed in a table, and once results are returned, they can be displayed in the histogram panel below. The page also supports batched evaluation, allowing multiple circuits to be selected, uploaded, and executed together to reduce cloud execution costs, which is particularly useful for iterative workloads with similar circuits such as VQE.

\subsection{NWQLib: Algorithm Solver Library}

\begin{figure}[h]
  \centering
\includegraphics[width=0.75\linewidth]{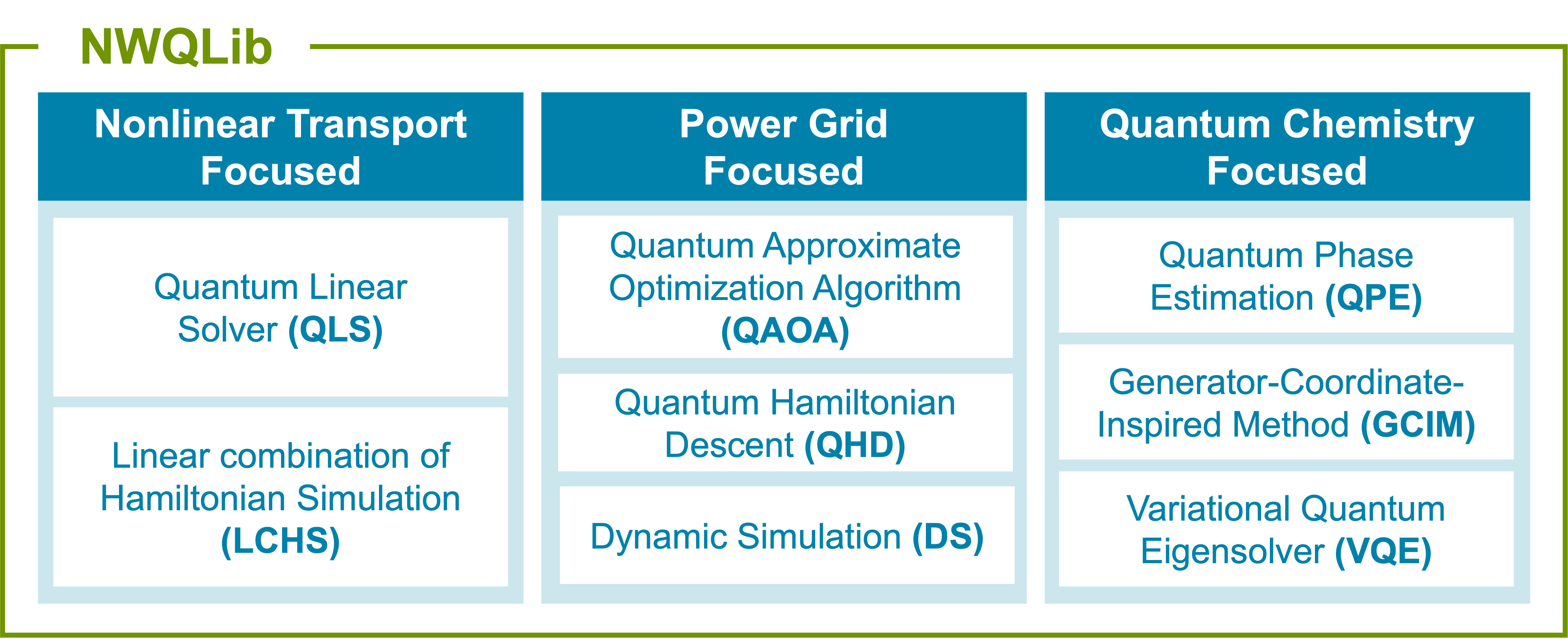}
  \caption{NWQLib algorithm solvers.}
  \label{fig:nwqlib}
\end{figure}

NWQLib is an algorithm solver library that has been co-designed with domain-specific computational scientists for gate-based quantum computing. The goal of NWQLib is to provide a general quantum solver abstraction layer for emerging domain applications. The current components are illustrated in Figure~\ref{fig:nwqlib}, and are primarily co-designed for applications in nonlinear transport (e.g., climate, fluid dynamics, earth system modeling, and biology), electric power grids, and quantum chemistry.

The solvers designed for nonlinear transport center around solving large linear and nonlinear systems for differential equations, see details in~\cite{li2025potential}. With that purpose, the solvers targeted include quantum linear solvers such as HHL~\cite{harrow2009quantum} and the emerging quantum simulation-based linear combination of Hamiltonian simulation (LCHS)~\cite{an2023linear}. The solvers for power grid center around quantum optimization, with interesting ones include Quantum Approximate Optimization Algorithm (QAOA)~\cite{farhi2014quantum, wang2024red}, Quantum Hamiltonian Descent~\cite{leng2023quantum}, and Dynamic Simulation~\cite{catli2025exponentially}. The solvers for quantum chemistry center around solving many-body electron systems, such as the ground state energy. The solvers targeted include VQE~\cite{peruzzo2014variational}, ADAPT-VQE~\cite{grimsley2019adaptive}, Generator-Coordinated Method (GCM)~\cite{zheng2024unleashed}, ADAPT-GCM~\cite{zheng2023quantum}, and Quantum Phase Estimation~\cite{nielsen2010quantum}. So far, our early evaluation of HHL for nonlinear transport and power grid can be found in~\cite{zheng2025early}. To accommodate for general applicability, the algorithm solver library should be developed in a broadly used quantum programming language such as Qiskit, with the ability to embed synthesized quantum circuits given configuration or invocation parameters. The quantum circuits are expected to be in QASM or NWQASM format for workflow integration.

\subsection{NWQASM Intermediate Representation}

NWQASM is based on, and fully compatible with, OpenQASM 2.0~\cite{cross2017open}, often referred to as QASM2 or QASM. To the best of our knowledge, QASM2 is the most widely used IR for quantum computing; nearly all quantum hardware vendors support QASM2 as (one of) their input formats. By building on QASM2, NWQASM therefore achieves the broadest applicability across quantum platforms. We are aware of OpenQASM 3.0~\cite{cross2022openqasm}, or QASM3. However, QASM3 more closely resembles a high-level programming language than an assembly-level IR. On the one hand, its control constructs are tightly coupled to IBM’s superconducting control system; on the other hand, its native advanced programming structures, such as loops and classical routines, are difficult to handle at runtime for workflow-level integration. Therefore, we retain QASM2 as the foundation and extend it with features that we believe are necessary for NWQWorkflow and for next-generation IR.

These extensions include:
\begin{itemize}
\item \textbf{Functional}: The present QASM was primarily designed for qubit-register-based quantum computing. To make it broadly applicable to next-generation quantum technologies, QASM should be extended to accommodate functional requirements from \emph{quantum memory}, \emph{quantum networking}, and \emph{quantum sensing}. In our prior work on quantum memory~\cite{liu2023quantum}, we proposed QASM extensions for abstracting and operating quantum random-access memory (QRAM) and random-access quantum memory (RAQM). In~\cite{wu2023qucomm}, we investigated collective communication and proposed QASM primitives for quantum networking during the design of a quantum networking processing unit (QNPU)~\cite{li2025quantum} or fault tolerant network interface card~\cite{guinn2023co}. In~\cite{jebraeilli2025stqs}, we designed QASM interfaces for spatial and temporal protocols for distributed quantum sensing. These new QASM abstractions and functionalities will be integrated into NWQASM.

\item \textbf{Timing}: During the development of noisy simulations for optimal protocol selection in entanglement purification~\cite{shi2025design}, we identified the need to introduce an explicit \emph{delay(qubit, cycles)} instruction, together with a global timing system, into QASM. This extension would enable more effective compiler scheduling and more accurate simulation of practical systems under time-dependent noise effects such as decay, emission, and relaxation.

\item \textbf{Storage}: When simulating very deep circuits, such as those arising in quantum simulations for nuclear systems~\cite{li2024deep} or in quantum phase estimation, a major limitation of QASM is its storage overhead. Program sizes can reach hundreds of megabytes, imposing a significant burden on data transfer and processing in compilers, runtimes, and simulators. At present, QASM programs are stored as plain text rather than in a compact binary representation. A native binary storage format should therefore be developed to store QASM programs more efficiently, analogous to how image formats such as JPEG are used for efficient image storage.
\end{itemize}

\subsection{Benchmarking}

\subsubsection{QASMBench}

For the NISQ scenario, we proposed QASMBench~\cite{li2023qasmbench} (\url{https://github.com/pnnl/QASMBench}). QASMBench is a QASM2-based benchmark suite designed to support the evaluation and characterization of NISQ-era quantum computing systems, tools, and simulators. It was introduced to address the lack of standardized, assembly-level benchmarks that can be used to systematically assess the performance of quantum hardware, compiler stacks, schedulers, and classical simulators. QASMBench consolidates commonly encountered quantum routines and kernels drawn from diverse domains such as quantum chemistry, simulation, linear algebra, searching and optimization, arithmetic, machine learning, fault tolerance, and cryptography. This coverage is intended to strike a practical balance between generality and usability for real device and simulator evaluations. QASMBench has already been widely used by the community. 

\subsubsection{FTCircuitBench}

For the FTQC scenario, we recently released FTCircuitBench~\cite{harkness2026ftcircuitbench} (\url{https://github.com/Adrian-Harkness/FTCircuitBench}). The motivation is to support the FTQC ecosystem by providing standardized circuits, compilation flows, and evaluation tools that span the full logical compilation stack. The motivation stems from the fact that large-scale, error-corrected quantum advantage will depend critically on fault-tolerant compilation and architecture, which impose different constraints and metrics than NISQ-era systems. The absence of vetted fault-tolerant benchmarks has made it difficult to compare optimization strategies and evaluate resource implications across compilation choices. FTCircuitBench incorporates a suite of representative quantum algorithm instances transpiled into the two major fault-tolerant logical computation models: \emph{Clifford+T} and \emph{Pauli Based Computation} (PBC). These include un-optimized baseline circuits and optimized logical circuits through our NWQEC compilers~\cite{wang2024optimizing, wang2025tableau} which are discussed later.

\subsection{Compilers}

\subsubsection{QASMTrans}
For the NISQ scenario, we developed and released QASMTrans compiler~\cite{hua2023qasmtrans} (\url{https://github.com/pnnl/qasmtrans}). QASMTrans is a device-aware, assembly-level quantum transpiler designed for NISQ devices, with a focus on scalability, runtime performance, and portability. It takes logical QASM2 as input and generates physical QASM2 according to a JSON-formatted device configuration file. This device configuration file specifies hardware characteristics such as qubit connectivity, basis gate sets, gate durations, and noise parameters. Later, in the NWQData subsection, we provide exemplar device JSON files for major existing quantum devices from multiple vendors. During transpilation, QASMTrans performs gate decomposition, qubit mapping, routing, and scheduling to generate a hardware-compliant circuit tailored to the target device. QASMTrans is implemented in C++ and is scalable to handle thousands of qubits and millions of gates. The current internal mapping logic assumes the IBM basis gate set (i.e., \texttt{U} and \texttt{CX}). Support for alternative basis gate sets on other devices is handled in the final compilation stage by decomposing and mapping \texttt{U} and \texttt{CX} gates to the target basis gates with lightweight fine-tuning. More advanced optimization can be achieved by directly optimizing and mapping circuits using the target device’s native basis gate set. Within the NWQWorkflow, QASMTrans serves as the primary NISQ compiler.

\subsubsection{NWQEC}
For the FTQC scenario, we developed the NWQEC compiler (\url{https://github.com/pnnl/nwqec}), targeting the \emph{Clifford+T} representation~\cite{wang2024optimizing} and the \emph{Pauli-Based Computation} (PBC) representation~\cite{wang2025tableau}. NWQEC is a fault-tolerant quantum circuit transpilation and optimization toolkit. Its core is implemented in C++, with a flexible Python interface for workflow integration. NWQEC takes logical QASM2 circuits as input, optimizes for minimal T-gate and Clifford gate counts, and generates FTQC circuits in both Clifford+T and PBC representations.

For the Clifford+T representation, the corresponding tool is TACO~\cite{wang2024optimizing}, which converts arbitrary logical circuits expressed in QASM2 into the Clifford+T gate set. Clifford+T is a standard basis for many fault-tolerant schemes, where T gates are expensive due to the overhead of magic-state distillation and lattice surgery. The conversion process employs high-precision synthesis algorithms (e.g., \emph{gridsynth}~\cite{ross2014optimal}) to express non-Clifford rotations using Clifford and T operations. This representation is essential for logical implementations in surface codes and related error-correcting codes that aim to minimize non-T resource costs.

For the PBC representation, the corresponding tool is TQC~\cite{wang2025tableau}, which translates logical QASM2 circuits into the PBC format. In the PBC model, computation is driven by a sequence of adaptive Pauli measurements and prepared \emph{magic} resource states. PBC can be viewed as a measurement-centric logical model that may reduce certain resource metrics, such as T-count or circuit depth, by re-expressing logical operations as sequences of Pauli measurements and rotations. TQC generates PBCs and applies Tfuse optimization and tableau-based passes to further reduce effective T-resource requirements and overall logical overhead.

\subsection{NWQControl}

\begin{figure}[h]
  \centering
\includegraphics[width=0.99\linewidth]{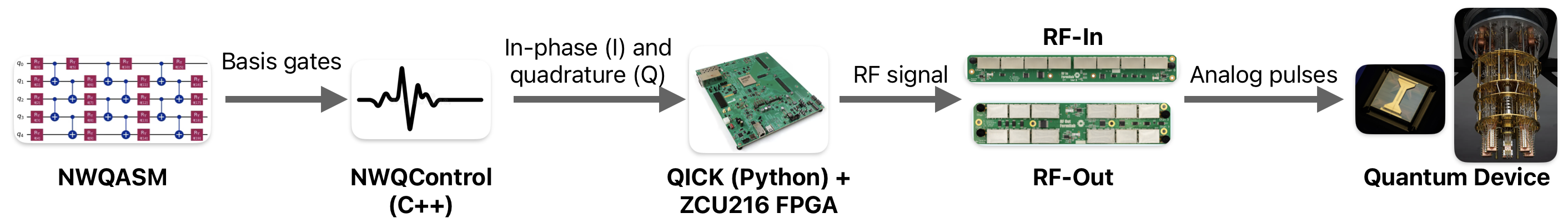}
  \caption{NWQControl Workflow.}
  \label{fig:nwqcontrol}
\end{figure}

The objective of NWQControl is to take physical QASM2 circuits generated by the compiler and produce optimized pulse sequences for quantum control. Pulse generation is typically tightly coupled to specific hardware platforms. To support an open and extensible ecosystem, NWQControl targets the Quantum Instrumentation Control Kit (QICK) framework~\cite{stefanazzi2022qick}. Designed for direct deployment on Xilinx RFSoC ZCU216 FPGA platforms, QICK provides high-speed DACs and ADCs with numerically controlled oscillators, mixers, and real-time pulse sequencers, accessible through a Python interface. Future work will explore integration with alternative control frameworks, such as  QubiC~\cite{xu2021qubic, xu2023qubic}.

NWQControl enables three key advances: (a) direct hardware integration through QICK for superconducting quantum testbeds, improving efficiency and avoiding the overhead of external toolchains; (b) internal validation through an integrated QuTiP~\cite{johansson2012qutip}-based simulator; and (c) optimization for quantum optimal control (QOC) by merging consecutive gate pulses to reduce latency and improve performance, synthesizing custom pulse sequences for frequent gate combinations along critical paths, as explored in our prior work~\cite{chen2023pulse}.

Figure~\ref{fig:nwqcontrol} illustrates the NWQControl workflow. A physical circuit in QASM or NWQASM format is provided as input to NWQControl, which is implemented in C++. The output of NWQControl consists of the in-phase and quadrature (I/Q) components, which represent sinusoidal (classical or microwave) signals using two orthogonal components corresponding to sine and cosine terms. These outputs are processed by QICK, which generates RF signals that are filtered and/or amplified through RF input/output daughter boards to deliver analog control pulses to the superconducting quantum processor housed within a dilution refrigerator. The FPGA-based NWQControl system can be connected to alternative classical accelerators, such as NVIDIA GPUs, through NVQLink~\cite{caldwell2025platform}, enabling low-latency responses for workloads including QEC decoding~\cite{wang2025fully, yin2024symbreak}, error mitigation~\cite{stein2023q, zheng2023bayesian}, and gradient-based optimization~\cite{stein2022eqc, l2024quantum}.

%NWQControl supports two pulse emission pathways to QICK hardware. In the first mode, NWQControl generates JSON-formatted pulse schedule files containing calibrated pulse sequences, which are subsequently parsed and executed using an included Python script. In the second mode, users invoke an \texttt{–emit} flag via the command-line interface, triggering direct pulse emission to the QPU through NWQControl’s PyBind-based API integration with QICK~\cite{stefanazzi2022qick}. This dual-mode architecture balances flexibility for offline analysis with the low-latency requirements of real-time quantum control.

\subsection{NWQSC Superconducting Quantum Testbed}

The primary focus of this whitepaper is the software workflow. However, to present a complete software–hardware co-design narrative, we briefly describe the construction of a superconducting testbed, referred to as NWQSC, that operates in conjunction with NWQWorkflow.

\begin{figure}[h]
  \centering
\includegraphics[width=0.99\linewidth]{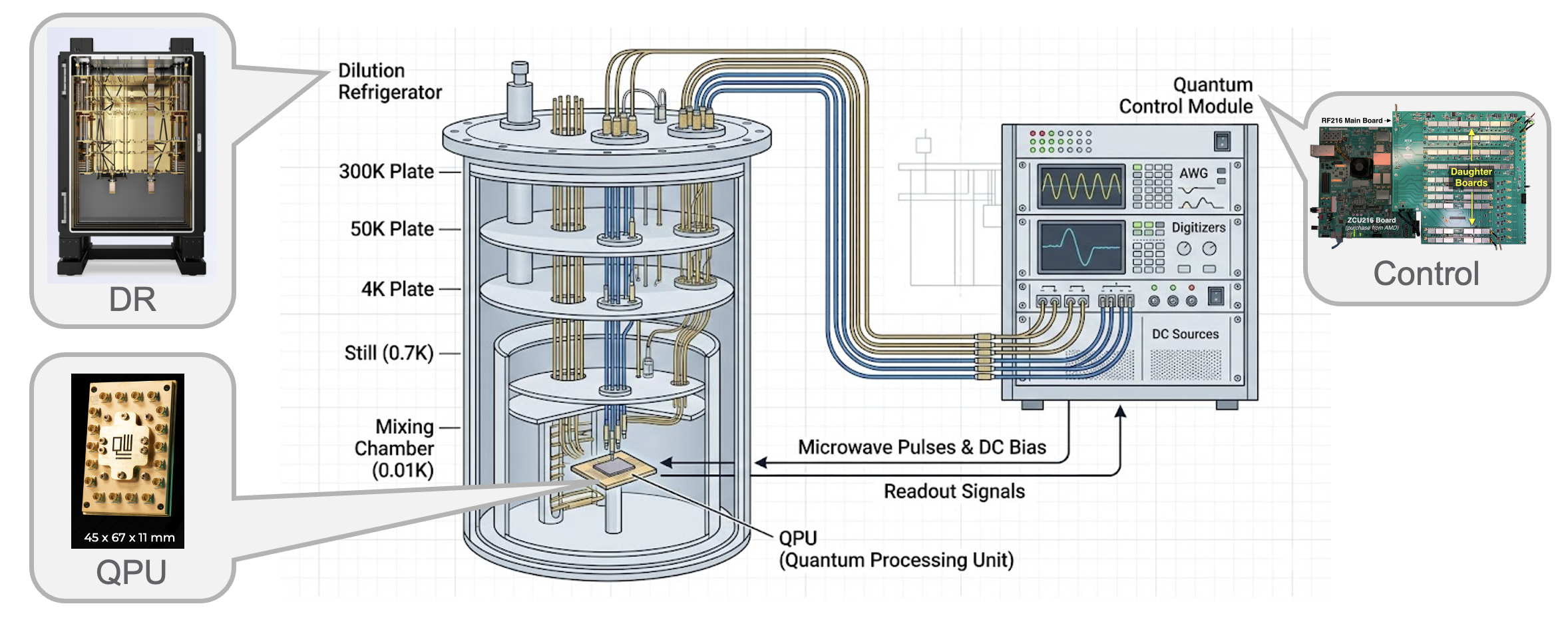}
  \caption{Superconducting Testbed.}
  \label{fig:nwqsc}
\end{figure}

As shown in Figure~\ref{fig:nwqsc}, a quantum testbed consists of three major components:
\begin{itemize}
\item \textbf{Control Module}: Commercial quantum control platforms are typically ASIC- or FPGA-based, such as product lines from Quantum Machines (QM) and Zurich Instruments (ZI). Community-developed solutions are predominantly FPGA-based, including QICK~\cite{stefanazzi2022qick} and QubiC~\cite{xu2021qubic, xu2023qubic}. For NWQSC, our control system is based on the QICK framework using RFSoC FPGAs in conjunction with NWQControl.
\item \textbf{Dilution Refrigerator (DR):} A dilution refrigerator progressively cools the system to near-zero temperatures to enable a superconducting environment. Major DR vendors include Bluefors, Oxford Instruments, Leiden Cryogenics, and Maybell Quantum.
\item \textbf{QPU:} The quantum processing unit (QPU) hosts the qubits, couplings, and resonators required for executing single- and two-qubit gates and for readout. Major superconducting QPU vendors include IBM, Google, IQM, Alice\&Bob, and QuantWare. Testbed QPUs are also fabricated in university and national research labs, such as the LPS Qubit Collaboratory (LQC).
\end{itemize}

\subsection{NWQSim Numerical Simulation on HPC}

\begin{figure}[h]
  \centering
\includegraphics[width=0.7\linewidth]{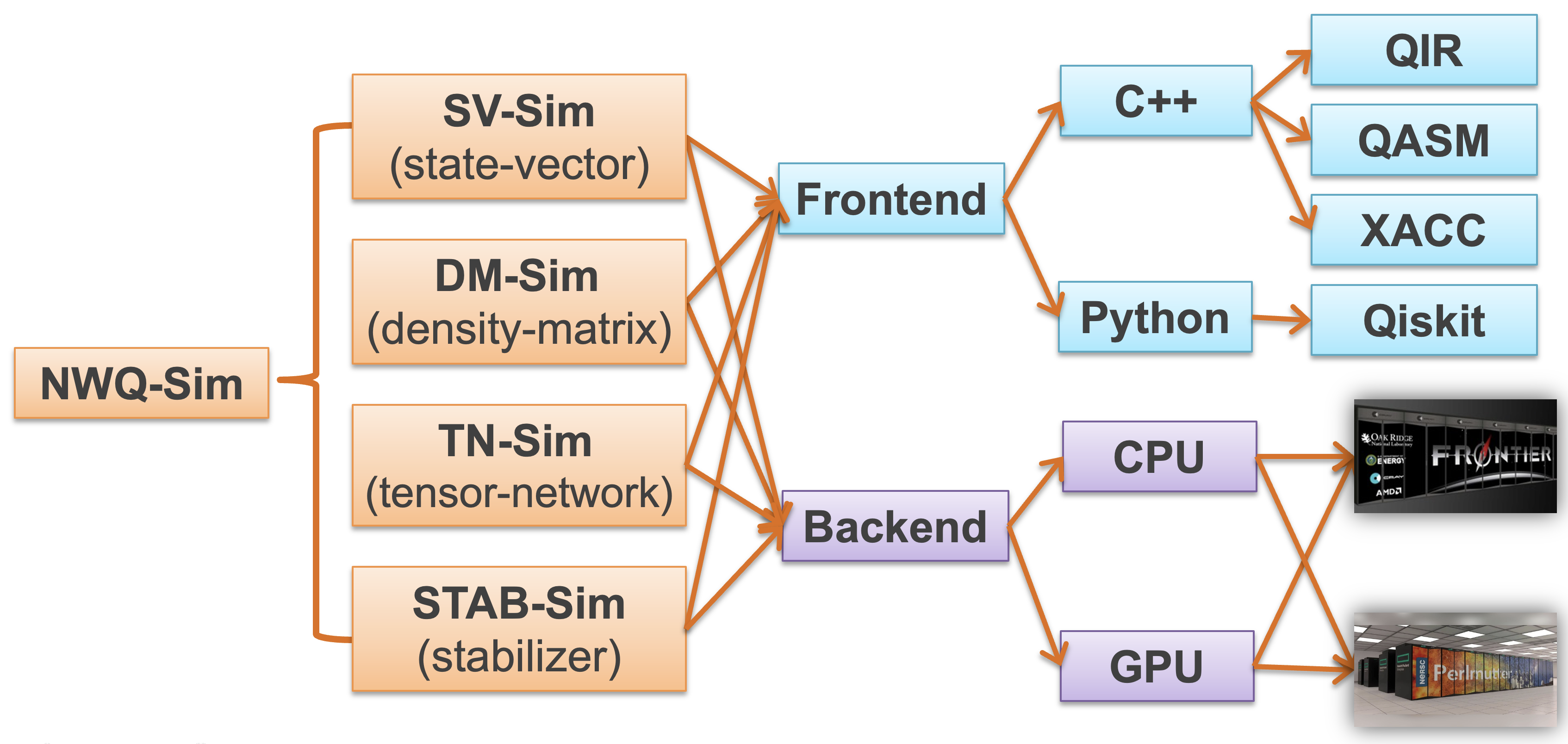}
  \caption{NWQSim components.}
  \label{fig:nwqsim}
\end{figure}

Numerical simulation of quantum circuits on classical HPC clusters remains crucial for validating quantum algorithms, investigating the impact of noise on quantum devices, and developing resilient quantum applications. In previous work, we developed a full-scale HPC quantum simulation capability, known as the NWQ-Sim package (\url{https://github.com/pnnl/nwq-sim}). As shown in Figure~\ref{fig:nwqsim}, NWQ-Sim comprises four simulator components: SV-Sim for state-vector functional simulation~\cite{li2021sv}, DM-Sim for density-matrix-based noisy simulation~\cite{li2020density}, TN-Sim for tensor-network-based low-rank simulation~\cite{hoyt2025tnsim}, and STAB-Sim for stabilizer simulation of Clifford circuits~\cite{garner2025stabsim}. These simulators accept OpenQASM~\cite{cross2017open}, QIR~\cite{li2024tanq}, and XACC~\cite{mccaskey2020xacc} as input through a C++ interface, and support Qiskit through a Python interface. The backends include both CPU- and GPU-based HPC systems, such as NERSC Perlmutter and OLCF Frontier. Most implementations (SV-Sim, DM-Sim, and STAB-Sim) are developed as native simulators without reliance on external libraries. The NWQ-Sim repositories provide comprehensive documentation for installation, configuration, and execution across a variety of system environments.

SV-Sim~\cite{li2021sv} (\url{https://github.com/pnnl/sv-sim}) is a state-vector-based, full-scale functional simulator for quantum circuit simulation on CPU- and GPU-based HPC systems. It supports execution on CPUs as well as NVIDIA and AMD GPUs. The design follows a shared-memory programming model, in which the large quantum state vector is distributed across, yet entirely stored within, GPU device memory. For scaling on HPC systems, SV-Sim relies on MPI and OpenSHMEM for multi-node CPU execution. For NVIDIA GPUs, it utilizes NVSHMEM to enable GPU-side direct communication, while for AMD GPUs, it employs MPI with RDMA or ROC\_SHMEM to achieve efficient inter-device communication. Aggressive gate fusion~\cite{li2024deep} is implemented to further improve performance. SV-Sim has been demonstrated to scale to 1,024 nodes with 4,096 NVIDIA A100 GPUs on the NERSC Perlmutter supercomputer for 42-qubit state-vector simulations, and to 4,096 nodes with 16,384 NVIDIA V100 GPUs on the OLCF Summit supercomputer for 40-qubit simulations.

\subsubsection{DM-Sim Density Matrix Simulation}

DM-Sim~\cite{li2020density} (\url{https://github.com/pnnl/dm-sim}) is a density-matrix-based, full-scale simulator capable of performing non-unitary, noisy quantum circuit simulations. In contrast to state-vector-based simulation, where memory usage scales as $2^n$ with n denoting the number of qubits, density-matrix simulation incurs a memory cost that scales as $4^n$. The computational cost of DM-Sim is therefore significantly higher than that of SV-Sim. Nevertheless, DM-Sim enables the simulation of realistic quantum systems by incorporating both coherent and incoherent noise processes. Its built-in noise models include depolarization, thermal relaxation, and readout noise~\cite{li2024tanq}. DM-Sim can be used to construct digital twins of real quantum devices, assisting quantum scientists in analyzing noise sources and their impact on algorithm performance. DM-Sim is designed with AVX-512 vectorized acceleration on CPUs, and tensor-core or matrix-core-based acceleration on NVIDIA and AMD GPUs~\cite{li2024tanq}. DM-Sim has been demonstrated to scale to 21-qubit density-matrix simulations on 1,024 nodes with 4,096 NVIDIA A100 GPUs on  Perlmutter. In our past work, DM-Sim has been used for investigating noise effect for quantum networking~\cite{ang2024arquin, shi2025design}, sensing~\cite{jebraeilli2025stqs}, and Ising simulation~\cite{li2024tanq}.

\subsubsection{TN-Sim Tensor Network Simulation}

TN-Sim~\cite{hoyt2025tnsim} (to be released at \url{https://github.com/pnnl/tn-sim} and \url{https://github.com/pnnl/nwq-sim}) is a tensor-network-based simulator for numerical quantum circuit simulation on HPC systems. It employs Matrix Product State (MPS) tensor network techniques built on the Tensor Algebra for Many-body Methods (TAMM)~\cite{mutlu2023tamm} framework and ITensor~\cite{fishman2022itensor} to enable distributed, HPC-scale tensor computations. Within TN-Sim, we developed a task-based parallelization scheme that supports parallelized gate contraction for wide quantum circuits with a large number of gates per circuit layer. TN-Sim is designed to support chemistry applications~\cite{bauman2025coupled}, error mitigation~\cite{stein2023q}, and circuit cutting~\cite{kan2024scalable}. 

\subsubsection{STAB-Sim Stabilizer Simulation}

STAB-Sim~\cite{garner2025stabsim} (to be released at \url{https://github.com/pnnl/stab-sim} and \url{https://github.com/pnnl/nwq-sim}) is a CPU- and GPU-accelerated stabilizer simulator designed for quantum error correction verification~\cite{garner2025exact}. It achieves significant speedups over existing CPU-based stabilizer simulators, such as the widely used Google Stim~\cite{gidney2021stim} and IBM Qiskit stabilizer simulator. STAB-Sim efficiently trivializes Clifford gate operations and exploits the massive parallelism available on modern GPUs, using warp- and wavefront-level primitives to accelerate the handling of measurement operations, which are typically performance bottlenecks. STAB-Sim also introduces a new error model~\cite{garner2025exact} that captures non-unitary effects in T1/T2 error channels with significantly improved performance while maintaining exact accuracy for most physical qubit systems. In~\cite{garner2025stabsim}, we demonstrate a chemistry application use case for measurement optimization~\cite{burns2024galic} and present a novel Clifford+T to PBC compilation optimization enabled by STAB-Sim.

\subsection{NWQData}

The goal of NWQData is to develop a standardized data format and to build a data repository for storing quantum metadata, including calibration, characterization, and benchmarking data for both testbed and commercial quantum devices, as well as runtime data generated from the simulation and execution of domain application circuits. These archived data can be leveraged for training AI models and for developing digital twins and heuristic error mitigation techniques, while also enabling tracking of development trends and the overall status of quantum computing and quantum error correction. To date,
\begin{itemize}
\item We have developed a quantum gate library (\url{https://github.com/pnnl/qasmtrans/tree/master/gatedef}) that collects commonly used single- and two-qubit gates. For each gate, we provide descriptions of its functionality and usage, along with the corresponding unitary matrix.
\item We have developed a device library (\url{https://github.com/pnnl/qasmtrans/tree/master/devicelib}) that aggregates quantum devices accessible through the OLCF Quantum Computing User Program (QCUP), Quantum Science Center (QSC) partners, and publicly available online repositories. This library covers devices from IBM, IonQ, IQM, OQC, Quafu, Quantinuum, and Rigetti.

\item We are building an execution data repository to archive data collected from our experimental and benchmarking studies~\cite{li2023qasmbench, stein2022eqc, bauman2025coupled, hua2022synergistic, jebraeilli2025stqs}, covering quantum devices from IBM, IonQ, Rigetti, and Quantinuum.
\end{itemize}

\section{Decade Outlook for NWQWorkflow}

\subsection{Short-term (1–2 years)}
In the short term (1–2 years), the primary objective would be to tightly integrate all components presented into a fully functioning pipeline for an initial end-to-end workflow demonstration. As discussed in Section~2.2 and illustrated in Figure\ref{fig:nwqflowchem}, a chemistry application, such as quantum simulation of benzene and free-base porphyrin (FBP) molecules, will be showcased through a domain-specific optimization toolchain (ExaChem$\rightarrow$SynGen$\rightarrow$TAMM$\rightarrow$Downfolding), the NWQWorkflow toolchain (NWQLib$\rightarrow$NWQASM$\rightarrow$ QASMTrans$\rightarrow$ NWQSim/NWQControl), and quantum testbeds (e.g., NWQSC, QCUP, and AQT). This effort will demonstrate the integration, configuration, and end-to-end functionality of the workflow on representative quantum testbeds under the NISQ setting.

\subsection{Mid-term (3–5 years)}
In the mid term (3–5 years), the primary objective is to transition from NISQ to FTQC. This transition is expected to be continuous and gradual rather than abrupt or disruptive. During this phase, hybrid NISQ/FTQC designs are anticipated to emerge, such as partial QEC for quantum simulation~\cite{akahoshi2024partially, dangwal2025variational} or simplified, iterative early-stage FTQC algorithms, such as iterative QPE~\cite{li2024iterative}, random-walk QPE~\cite{granade2022using}, low-depth QPE~\cite{ni2023low}, variational QPE~\cite{klymko2022real}, etc. The NWQWorkflow should remain flexible and adaptable to support these hybrid execution models.

\subsection{Long-term (5–10 years)}
In the long term (5–10 years), the objective is to demonstrate domain applications using fully fault-tolerant quantum computing algorithms, such as Quantum Phase Estimation (QPE), on fault-tolerant quantum hardware. A key transition will involve replacing the NISQ components of the workflow with their fault tolerant counterparts, along with integrating QEC protocols into NWQControl and the underlying hardware stack. The chemistry workflow in Figure~\ref{fig:nwqflowchem} or those identified in our recent workshop~\cite{alexeev2025perspective} and perspective papers~\cite{alexeev2024quantum, buchs2025role} can be demonstrated under the FTQC setting.

\section{Conclusion}
This whitepaper presents NWQWorkflow, an end-to-end, full-stack software–hardware toolchain designed to map emerging quantum domain applications onto both NISQ and fault-tolerant quantum computing (FTQC) hardware. The workflow encompasses the programming environment, intermediate representations, compilers, benchmarking tools, HPC simulators, control systems, data collection, and quantum testbeds. NWQWorkflow is expected to serve as an integrated toolset for fostering collaborative efforts within the quantum computing community and for facilitating the transition of quantum computing from conceptual exploration to practical utility.

\section*{Acknowledgment}

\textbf{Contribution}: I would like to thank the following people for the significant contributions to the workflow: Meng Wang (PNNL \& UBC), Samuel Stein (PNNL), Chenxu Liu (PNNL),  Muqing Zheng (PNNL), Yanfei Li (PNNL), Sean Garner (PNNL \& UW), Aaron Hoyt (PNNL \& UW), Fei Hua (PNNL \& Rutgers), Matthew Burns (PNNL \& URochester), Anastashia Jebraeilli (PNNL \& UGA), Peiyi Li (PNNL \& NCSU), Nicholas Bauman (PNNL), Bo Peng (PNNL), Shuwen Kan (PNNL \& Fordham), Ming Wang (PNNL \& NCSU), Adrian Harkness (PNNL \& Lehigh), Yue Shi (PNNL \& UW), Anbang Wu (PNNL \& UCSB), Chunshu Wu (PNNL), and Drew Rebar (PNNL). 

\vspace{4pt}\noindent\textbf{Support:} Special thanks to Wendy Shaw (PNNL), Travis Humble (ORNL), Robert Rallo (PNNL), Karol Kowalski (PNNL), Lou Terminello (PNNL), James Ang (PNNL), Marvin Warner (PNNL), Cindy Bruckner-Lea (PNNL), Courtney Corley (PNNL), Sriram Krishnamoorthy (PNNL), Nathan Baker (PNNL), Karl Mueller (PNNL), Edmond Hui (PNNL), Mark Raugas (PNNL), Gail Anderson (PNNL), Kevin Barker (PNNL), Michael Spradling (PNNL), Nathan Wiebe (PNNL\&U Toronto), Yufei Ding (UCSD), Kai-Mei Fu (UW), Katherine Klymko (LBL), Norman Tubman (NASA), Ying Mao (Fordham), In-Saeng Suh (ORNL), Gilles Buchs (ORNL), Andrew Sornborge (LANL), Shinjae Yoo (BNL), Sara Sussman (FermiLab), Yuri Alexeev (ANL), Ji Liu (ANL), Steven Girvin (Yale), Charles Black (BNL), Andrew Houck (Princeton), Andrew Cross (IBM), David Mckay (IBM), Mark Ritter (IBM), CJ Newburn (NVIDIA), Jim Dinan (NVIDIA), Jin-Sung Kim (NVIDIA), Brandon Potter (AMD), Martin Roetteler (Microsoft), Bettina Heim (Microsoft), Erica Stump (IonQ), Frank Mueller (NCSU), Prashant Nair (UBC), Eddy Zhang (Rutgers), Huiyang Zhou (NCSU) for the tremendous support to our work. 

\vspace{4pt}\noindent\textbf{Funding:} Within NWQWorkflow, NWQSim, NWQEC, NWQASM and NWQStudio were mainly supported by the U.S. Department of Energy, Office of Science, National Quantum Information Science Research Centers, Quantum Science Center (QSC). NWQLib, NWQControl, NWQSC were supported by the Quantum Algorithms and Architecture for Domain Science Initiative (QuAADS), under the Laboratory Directed Research and Development (LDRD) Program at Pacific Northwest National Laboratory (PNNL). QASMBench and QASMTrans were mainly supported by the U.S. Department of Energy, Office of Science, National Quantum Information Science Research Centers, Co-design Center for Quantum Advantage (C2QA) under contract number DE-SC0012704, (Basic Energy Sciences, PNNL FWP 76274). The chemistry part and integration were supported by the Embedding QC into Many-body Frameworks for Strongly Correlated Molecular and Materials Systems project, which is funded by the U.S. Department of Energy, Office of Science, Office of Basic Energy Sciences (BES), the Division of Chemical Sciences, Geosciences, and Biosciences (under award 72689). The initial stage NWQSim was supported by PNNL’s
Quantum Algorithms, Software, and Architectures (QUASAR)
LDRD Initiative. The AI integration was supported by the U.S. Department of Energy, Office of Science, the Transformational AI Models Consortium (ModCon), Quantum Seeding project (PNNL FWP 86646). 

\vspace{4pt}\noindent\textbf{Resources:} This research used resources of the Oak Ridge Leadership Computing Facility (OLCF), which is a DOE Office of Science User Facility supported under Contract DE-AC05-00OR22725. This research used resources of the National Energy Research Scientific Computing Center (NERSC), a U.S. Department of Energy Office of Science User Facility located at Lawrence Berkeley National Laboratory, operated under Contract No.DE-AC02-05CH11231. This research used resources of Research Computing (RC) at PNNL. The Pacific Northwest National Laboratory is operated by Battelle for the U.S. Department of Energy under Contract DE-AC05-76RL01830.

\bibliographystyle{unsrt}
\bibliography{ref}

\end{document}